\documentclass{ws-procs975x65}
\usepackage{hyperref}
\def\beq{\begin{equation}}
\def\eeq{\end{equation}}
\begin{document}

\title{Solutions of gravitational field equations for\\ weakly birefringent spacetimes \\{\normalfont\small\textit{Oral contribution to the session on Constructive Gravity\\ 15th Marcel Grossmann Meeting 2018 in Rome}}}
\author{Nils Alex}

\address{Department Physik, Friedrich-Alexander-Universit\"at Erlangen-N\"urnberg,\\
Staudtstr.~7, 91058 Erlangen, Germany\\
E-mail: nils.alex@fau.de}

\begin{abstract}
  The constructive gravity programme applied to electrodynamics with vacuum birefringence yields the---up to unknown gravitational constants---unique compatible gravity theory for the underlying non-metric geometry. Starting from a perturbative variant of this procedure, we solve the resulting gravitational field equations for the geometry perturbation and point out the non-metric refinements to the linear Schwarzschild and the gravitational wave solution.
\end{abstract}


\bodymatter

\section{Introduction}

Equations of motion for a matter field (e.g.~an electromagnetic potential) are always \emph{incomplete} in the sense that the coefficients in these equations, which are functions of the geometric field to which the matter field couples, are not determined. General relativity, in its essence, \emph{closes} Maxwell's equations by providing dynamic equations for the metric tensor. The gravitational closure programme provides a prescription for performing this closure for any kind of matter theory subject to a few conditions.\cite{D_ll_2018} Applying a perturbative variant of this procedure to a certain generalisation of Maxwell electrodynamics which does not exclude vacuum birefringence \emph{a priori} has been shown to yield gravitational field equations that predict the dynamics of these birefringent perturbations.\cite{Schneider_2017}

\section{Birefringent electrodynamics}

The most general theory for an electrodynamic potential has been derived by Hehl and Obukhov in Ref.~\refcite{Hehl_2003} from the principles of conservation of electric charge and magnetic flux and a linear constitutive law. Instead of the usual metric tensor $g$ from Maxwell electrodynamics, the potential couples to a rank 4 tensor field $G$ with symmetries
\beq
  G^{abcd} = G^{cdab} = -G^{bacd}
\eeq
via the action
\beq
  S[A,G) = \int \mathrm d^4x\,\omega_G\, G^{abcd} F_{ab} F_{cd},
\eeq
where $F = \mathrm dA$ denotes the electromagnetic field strength and $\omega_G$ a 1-density built from $G$. This theory is called \emph{General Linear Electrodynamics (GLED)}. Maxwell electrodynamics is recovered by setting $G^{abcd} = g^{ac} g^{bd} - g^{ad} g^{bc} - \sqrt{-g}\,\epsilon^{abcd}$ and $\omega_G = \frac{1}{24} \epsilon_{abcd} G^{abcd} = \sqrt{-g}$.

A very distinctive property of GLED is the causality of the field equations: While wave covectors in the WKB limit of Maxwell electrodynamics are selected by the causality condition $0 = P(k) = g(k,k)$, which is a homogeneous polynomial of degree two, the causality condition of GLED is of degree four.\cite{Hehl_2003} This allows for more roots of $P$ which introduces the possibility of a polarisation-dependent speed of light---birefringence in vacuo.

\section{Gravitational closure of weakly birefringent electrodynamics}

While the gravitational closure procedure is fully laid out, it remains very difficult to follow through because the problem boils down to a set of infinitely many homogeneous first-order PDEs. There is, however, a way around this obstacle: \emph{Perturbative} gravitational closure, which is only concerned with gravitational dynamics of small perturbations of the geometric field. This is also the sector of great phenomenological relevance, because corrections to Einstein gravity and field theories coupling to the geometry should---if present at all---be rather minute.

In Ref.~\refcite{Schneider_2017}, the perturbative gravitational closure procedure has been applied to perturbations $H$ of the GLED geometry $G$,
\beq
  G^{abcd} = \eta^{ac} \eta^{bd} - \eta^{ad} \eta^{bc} - \epsilon^{abcd} + H^{abcd},
\eeq
which results in a 11-parameter family of field equations for the perturbation $H$, where $H$ has been gauge-fixed and decomposed according to a $3+1$ split into a set of spatial fields (Greek indices being spatial indices from 1 to 3, raised and lowered at will with the positive definite spatial metric $\gamma$):

\vspace{2ex}
\noindent\textbf{transverse traceless (tt) tensor modes} $U_{\alpha\beta}, V_{\alpha\beta}, W_{\alpha\beta}$ satisfying
\beq
\begin{aligned}
  0 = U_{\lbrack\alpha\beta\rbrack},\quad 0 = \partial_\alpha U^{\alpha\beta} \quad \text{and} \quad 0 = \gamma_{\alpha\beta} U^{\alpha\beta},\\
  V_{\alpha\beta}, W_{\alpha\beta} \quad \text{similar},
\end{aligned}
\eeq
\noindent\textbf{solenoidal vector modes} $U_\alpha, W_\alpha, B_\alpha$ satisfying
\beq
  0 = \partial_\alpha U^\alpha, \quad W_\alpha, B_\alpha \quad \text{similar},
\eeq
\noindent\textbf{scalar modes} $V,W,\widetilde U, \widetilde V, A$.
\vspace{2ex}

Einstein gravity corresponds to the sector where only $U_{\alpha\beta}, B_\alpha, \widetilde U,$ and $A$ are non-zero. In the following sections, we will solve the field equations for two prototypical matter distributions: The inertial point mass, which yields corrections to the linear Schwarzschild solution of Einstein gravity, and the vacuum, which reveals the causality of the theory.

\section{Inertial point mass source}
An inertial point mass resting at the spatial origin of a coordinate chart is described by the world line
\beq
  \gamma^a(\tau) = \tau \delta^a_0
  \label{point_particle}
\eeq
and produces a contribution of $-M\delta^{(3)}(x)$ in the equation of motion which corresponds to the variation of the action with respect to $A$. Since the matter distribution (\ref{point_particle}) is stationary, we may restrict our attention to stationary solutions for the gravitational field by setting all time derivatives to zero. In this setup, the field equations for tt and solenoidal vector modes are homogeneous and yield
\begin{align}
  0 &= T_{\alpha\beta} \quad \text{for all tt modes}, \\
  0 &= V_{\alpha} \quad \text{for all solenoidal vector modes}.
\end{align}

The field equations for the scalar modes read
\beq
\begin{aligned}
  0 &= \sigma_1 \Delta V + \sigma_2 \widetilde U + \sigma_4 \widetilde V - 4 \sigma_2 A, \\
  0 &= \sigma_4 \Delta V + \sigma_5 V + \sigma_6 \Delta W + \sigma_7 W + (-3\sigma_1 + \sigma_2) \widetilde U + (-9\sigma_1 + 3 \sigma_2) \widetilde V\\
  &\hphantom{= } + (-12\sigma_1 + \sigma_2) A, \\
  0 &= \sigma_6 \Delta V + \sigma_7 V + \sigma_8 \Delta W - \sigma_5 W, \\
  0 &= (-2\sigma_1 + \frac{2}{3} \sigma_2 - \frac{2}{9}\sigma_3) \Delta\Delta V - \frac{2}{3} \sigma_3 \Delta \widetilde U + \sigma_9 \Delta \widetilde V + \sigma_{10} \widetilde V \\
    &\hphantom{= } + (4 \sigma_2 + \frac{4}{3} \sigma_3) \Delta A, \\
  -M &= (-4 \sigma_1 + \frac{8}{3}\sigma_2) \Delta\Delta V + 4 \sigma_2 \Delta \widetilde U + (6\sigma_2 + 2 \sigma_3) \Delta \widetilde V, 
\end{aligned}
\eeq
where the constants $\sigma_i$ denote combinations of the 11 independent gravitational constants. The general solution of this system admits the form
\begin{align}
  V(x) &= \frac{M}{4\pi r} \lbrack G_1 + G_2 f_\nu(r) + G_3 f_\rho(r)\rbrack\nonumber\\
  W(x) &= \frac{M}{4\pi r} \lbrack G_4 + G_5 f_\nu(r) + G_6 f_\rho(r)\rbrack\nonumber\\
  \widetilde U(x) &= \frac{M}{4\pi r} \lbrack G_7 + G_8 f_\mu(r) + G_9 f_\nu(r) + G_{10} f_\rho(r) \rbrack\label{stationary_sols}\\
  \widetilde V(x) &= \frac{M}{4\pi r} \lbrack G_{11} f_\mu(r)\rbrack\nonumber\\
  A(x) &= \frac{M}{4\pi r} \lbrack G_{12} + G_{13} f_\mu(r) + G_{14} f_\nu(r) + G_{15} f_\rho(r)\rbrack\nonumber
\end{align}
with functions
\beq
  f_\alpha(r) = \begin{cases} \cos(\sqrt{-\alpha}r) & \alpha < 0 \\ \mathrm e^{-\sqrt{\alpha}r} & \alpha \geq 0\end{cases}
\eeq
and constants $G_i, \mu, \nu, \rho$ depending on the 11 gravitational constants.\footnote{This is the result for one of two cases which have to be considered, depending on the exact values of the gravitational constants. The solution in the other case contains products of exponential and trigonometric functions of $r$, but the discussion below and especially the result (\ref{stationary_reduced}) remain the same.}

The solutions (\ref{stationary_sols}) can be regarded as corrections to the linearized Schwarzschild solution of Einstein gravity, which in the context of GLED reads
\beq
  V = 0, \quad W = 0, \quad \widetilde V = 0, \quad \widetilde U(x) = 4 \kappa \frac{M}{4\pi r}, \quad \text{and} \quad A(x) = \kappa \frac{M}{4\pi r}.
\eeq
If (\ref{stationary_sols}) shall only introduce \emph{short-ranging} Yukawa corrections---instead of \emph{long-ranging} Coulomb corrections---, the constants $G_1$ and $G_4$ must vanish and the ratio $G_7 / G_{12}$ must be equal to 4. These conditions are satisfied by only one condition on the 11 independent gravitational constants and greatly simplify the form of (\ref{stationary_sols}) to
\beq
\begin{aligned}
  V(x) &= 0, \\
  W(x) &= 0, \\
  \widetilde U(x) &= \frac{M}{4\pi r} \lbrack \alpha + \beta \mathrm e^{-\sqrt{\mu}r} \rbrack, \\
  \widetilde V(x) &= \frac{M}{4\pi r} \gamma \mathrm e^{-\sqrt{\mu}r}, \\
  A(x) &= \frac{M}{4\pi r} \lbrack \frac{1}{4} \alpha - \frac{1}{4} (\beta + 3\gamma) \mathrm e^{-\sqrt{\mu}r} \rbrack. 
\end{aligned}
\label{stationary_reduced}
\eeq

Arriving at (\ref{stationary_reduced}) concludes---perturbatively and only for this special matter distribution---the gravitational closure programme for a theory of electrodynamics with vacuum birefringence. The result is a \emph{prediction} of where and how non-metric refinements are produced by an inertial point mass, depending on 4 unknown gravitational constants. The ramifications of such refinements have been investigated both on the astronomic scale for e.g.~galaxy rotation curves\cite{Rieser_2020} as well as on the quantum scale for quantum electrodynamics\cite{GrosseHolz_2017}. Experimental tests based upon these studies could now, in principle, provide approximations or bounds for the unknown parameters in (\ref{stationary_reduced}).

\section{Vacuum solutions}

In the case of a spacetime with no matter content, the gravitational field equations are a system of coupled homogeneous wave equations with mass and damping terms. A very critical property of the solutions of this system can be examined by calculating the complex eigenvalues of the time evolution: the stability of solutions with respect to small variations of the initial data. Stability in this sense is guaranteed if and only if the real part of each eigenvalue is non-positive. Because we are interested in investigating the sector of the theory which produces only small phenomenological refinements to Einstein gravity, we implement this stability conditions and find that the remaining 10 gravitational constants must satisfy two additional conditions.

Furthermore, for propagating gravitational fields it is desirable that the causality condition in the WKB limit coincides with the Minkowski causality condition $\eta(k,k) = 0$. Otherwise, consistent co-evolution of gravitational and electrodynamic fields as proposed by the gravitational closure programme would not be possible. This imposes one further condition on the gravitational constants. With this final condition implemented, we arrive at a 7-parameter family of gravitational field equations which in vacuo simplify to

\vspace{2ex}
\noindent\textbf{one massless wave equation}
\beq
  0 = \Box U_{\alpha\beta},
\eeq
\noindent\textbf{massive wave equations}
\beq
\begin{aligned}
  0 &= \Box V_{\alpha\beta} + m V_{\alpha\beta} = \Box W_{\alpha\beta} + m W_{\alpha\beta}, \\
  0 &= \Box U_\alpha + m U_\alpha = \Box W_\alpha + m W_\alpha, \\
  0 &= \Box V + m V = \Box W + m W, \\
  0 &= \Box \widetilde V + m_\text{S} \widetilde V,
\end{aligned}
\eeq
\noindent\textbf{constraint equations}
\beq
\begin{aligned}
  B_\alpha &= -\frac{1}{2} \dot U_\alpha, \\
  \widetilde U &= -\frac{1}{3} \Delta V - \frac{\sigma + 3}{2} \widetilde V, \\
  A &= \frac{1}{4} \ddot V + \frac{1}{4} \frac{\sigma - 3}{2} \widetilde V,
\end{aligned}
\eeq
with three combinations of gravitational constants: a mass $m$, a second mass $m_\text{S}$ for the scalar mode $\widetilde V$, and $\sigma$. Comparing this result with the vacuum equation for Einstein gravity,
\beq
  0 = \Box U_{\alpha\beta} \quad \text{+ constraint equations},
\eeq
we immediately see that the metrically inducible mode $U_{\alpha\beta}$ still obeys a \emph{massless} wave equation, while all non-metric modes evolve according to \emph{massive} wave equations. This fact has a great impact on the phenomenology of the non-metric corrections: In e.g.~binary systems, massive modes are only produced above a certain threshold\cite{M_ller_2018} and disperse during their propagation. That is, both the \emph{generation} as well as the \emph{detection} of these modes will be suppressed, depending on the exact values of the gravitational constants.

It is remarkable that, in addition to the one metric and two non-metric tt modes, also \emph{solenoidal vector modes} and \emph{scalar modes} propagate. This introduces new ways a gravitational wave can affect test matter, which is seen by calculating the geodesic deviation of dust spheres: a tt mode propagating in $z$-direction deforms geodesic spheres in $x$- and $y$-direction in a volume preserving way. A trace-free scalar mode propagating in the same direction, on the other hand, would lead to volume-preserving oscillations in $z$- and both lateral directions---exhibiting qualitatively completely new behaviour, which is quantitatively suppressed by the unknown mass. Pure trace modes also introduce a new quality as they make the whole volume of a geodesic sphere oscillate.

\emph{How} these modes can be generated has been studied in Ref.~\refcite{M_ller_2018} for non-gravitationally bound systems, because emission of gravitational waves from a gravitationally bound system is a second order effect which requires knowledge of the gravitational theory to this order. Currently, two different approaches (for one approach, see Ref.~\refcite{Alex_2020}) for obtaining the second order of the gravitational theory consistent with GLED are being pursued. Their results will fill this gap and eventually allow to answer the question: What would the non-metric refinements to emission, propagation and detection of gravitational waves be if the vacuum was birefringent?


\end{document}